\documentclass[letterpaper, 10 pt, conference]{ieeeconf}  % Comment this line out if you need a4paper

\IEEEoverridecommandlockouts                              % This command is only needed if 
\usepackage{graphics} % for pdf, bitmapped graphics files
\usepackage{epsfig} % for postscript graphics files
\usepackage{mathptmx} % assumes new font selection scheme installed
\usepackage{times} % assumes new font selection scheme installed
\usepackage{amsmath} % assumes amsmath package installed
\usepackage{amssymb}  % assumes amsmath package installed
\usepackage{mathtools}
\title{\LARGE \bf
Fast and Efficient Estimation of Resonant Modes: A Case Study of  Mechanical Drivelines*
}

\author{Amin Rezaeizadeh$^{1}$ and Silvia Mastellone$^{1}$% <-this % stops a space
\thanks{*This work was supported by NCCR Automation and SNSF}% <-this % stops a space
\thanks{$^{1}$Amin Rezaeizadeh and Silvia Mastellone are with the Institute of Electric Power Systems, University of Applied Science Northwest Switzerland, Windisch-Brugg, Switzerland. 
        {\tt\small amin.rezaeizadeh@fhnw.ch, silvia.mastellone@fhnw.ch}}%
}

\begin{document}

\maketitle
\thispagestyle{empty}
\pagestyle{empty}

%%%%%%%%%%%%%%%%%%%%%%%%%%%%%%%%%%%%%%%%%%%%%%%%%%%%%%%%%%%%%%%%%%%%%%%%%%%%%%%%
\begin{abstract}

This work presents the development of an online parameter estimation algorithm for the identification
of resonating modes in a linear system of arbitrary order. 
The method employs  a short-time
Fourier transform of the input and output signals and uses a recursive least square (RLS) algorithm to detect resonant frequencies and damping factors of the resonant modes. 
\end{abstract}

%%%%%%%%%%%%%%%%%%%%%%%%%%%%%%%%%%%%%%%%%%%%%%%%%%%%%%%%%%%%%%%%%%%%%%%%%%%%%%%%
%\section{INTRODUCTION}
% Current/frequency converters (Figure 1.1) are a vital component in various industries, including
% but not limited to heating, ventilation, air cooling (HVAC) systems, oil and gas, marine, and
% mining industries. These converters play a crucial role in converting electrical power to mechanical
% power and performing the reverse conversion in power generators. However, the control
% engineers responsible for the control structure and parametrization of the converters often lack
% a comprehensive understanding of the complex mechanical processes involved in these various
% industries. 

\section{Problem Statement}
Consider a SISO linear time-invariant system with resonant modes described by the  frequency response  $G(j\omega)$. The input and output signals are related through the following equation
\begin{equation}
    Y(j\omega)=G(j\omega)U(j\omega),
\end{equation}
where $U(j\omega)$ and $Y(j\omega)$ are, respectively, the discrete Fourier transform (DFT) of the input and output signals.

Our  objective is to determine the damping factors and resonant frequencies of the system  by identifying  the resonating poles of $G$. This task can be simplified by assuming that around each resonant frequency, the behavior of the system is mainly dominated by the complex-conjugate poles. That is, $G(j\omega)$ can be approximated by 
\begin{equation}
    G(j\omega)\approx \frac{c}{-\omega^2+2j\alpha \omega +\beta},
\end{equation}
where $c$ is a constant complex number representing the effect of other terms in the frequency response at frequency $\omega$.  The parameters $\alpha$ and $\beta$  depend on the resonant frequency and the damping factor.  

The resonant frequency, $f_r$, and the damping factor, $\zeta$, are then readily calculated by
\begin{equation}\label{eq:fr}
    f_r=\frac{1}{2\pi}\sqrt{\beta-\alpha^2},~~ \zeta=\frac{\alpha}{\sqrt{\beta}}.
\end{equation}
Plugging in the reduced order model of $G$, and splitting the signals into real and imaginary parts, lead to 
%\begin{equation}
%\begin{split}
\begin{multline}    \label{eq:main}
            \left(-\omega^2+2j\alpha \omega +\beta\right)\left(Y_r(\omega)+jY_i(\omega)\right)=\\
    (c_{r}+jc_{i})\left(U_r(\omega)+jU_i(\omega)\right),
\end{multline}
where $Y_r$, $Y_i$, $U_r$ and $U_i$ are, respectively, the real and imaginary parts of the output and the input DFTs at frequency $\omega$. 

Eq. \ref{eq:main} can be split into real and imaginary parts and reformulated as follows
\begin{equation}
\underbrace{ \omega^2\begin{bmatrix} Y_r  \\ Y_i \end{bmatrix}}_\text{$z(\omega)$}
   =
    \underbrace{\begin{bmatrix}
        -2\omega Y_i & Y_r & -U_r & U_i \\
       2\omega Y_r & Y_i & -U_i & -U_r 
    \end{bmatrix}}_\text{$H(\omega)$}x,
\label{eq:zHx}
\end{equation}
where $x$ is the parameter vector defined as 
\begin{equation*}
x:=\begin{bmatrix} \alpha& \beta& c_{r}& c_{i}  \end{bmatrix}^T.
  \end{equation*}

The Eq. \ref{eq:zHx} is in the form of $z(\omega)=H(\omega)x$, where $z$ and $H$ both depend on the frequency, whereas $x$ is a constant parameter vector. Therefore, the parameter vector, $x$, can be estimated using a computationally efficient recursive least square (RLS) approach. The RLS algorithm is only applied  only around the resonant frequency (for example within 20\% bound of the initial guess), and recursively updates the estimated parameter $\hat{x}$  according to the following set of equations,
\begin{equation}
    \hat{x}_{n+1}=\hat{x}_n + K(z(\omega_n)-H(\omega_n)\hat{x}_n),
\end{equation}
where subscript $n$ captures the index of the DFT frequency points used for the estimation, and the matrix $K$ is the optimal gain matrix derived by the following equation
\begin{equation}
    K=PH^T(HPH^T+R)^{-1},
\end{equation}
where $R$ is the covariance matrix on the $z(\omega)$, and $P$ is the covariance matrix of the estimate given by,
\begin{equation}
    P=(I-KH)P(I-KH)^T+KRK^T.
\end{equation}

\begin{figure}[thpb]
      \centering
      
 \includegraphics[scale=.5]{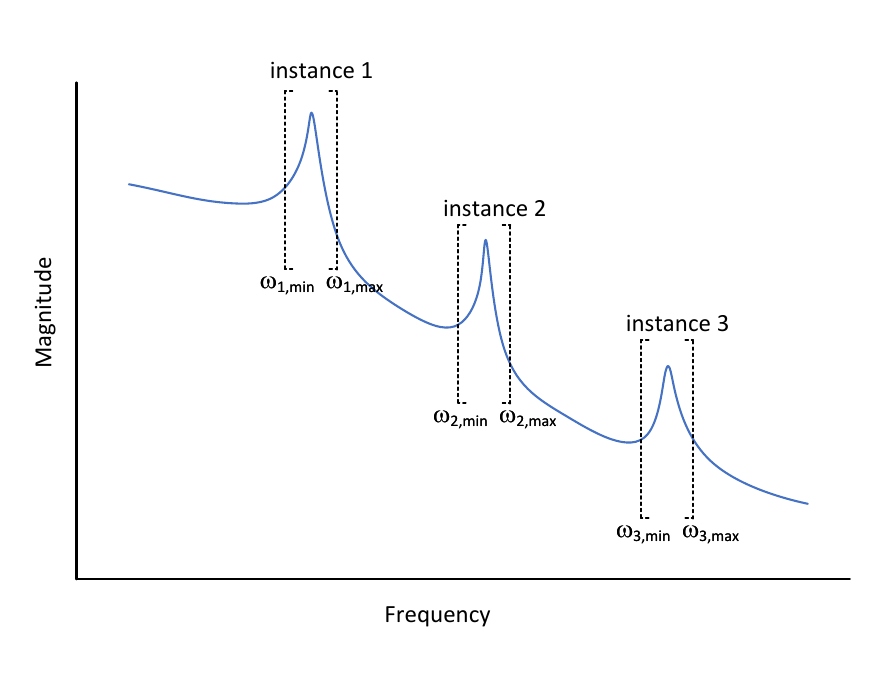}    
      \caption{online estimation of multiple resonant modes in parallel instances.   }
      \label{fig:instances}
   \end{figure}

The resonant frequency and the damping factor are then computed online according to Eq. \ref{eq:fr}. 
For the case where multiple resonant modes are required to be estimated, the estimation schemes can be run in parallel instances (see Fig. \ref{fig:instances}), independent of one another, if the modes are distant.

\section{Case study: Mechanical driveline torsional vibration}
Torsional vibrations in mechanical rotating  drivelines cause gear wear that eventually leads to reduced performance at best and even shaft breakage and  system failure. To avoid these issues, it is necessary to analyze  the torsional response characteristics of the system to diagnose malfunctioning and employ preventive measures. The magnitude of torsional vibrations is influenced by the amount of torsional excitation and the gap between the excitation frequencies and natural frequencies. To prevent torsional excitation frequencies from aligning with natural torsional frequencies, measuring the natural torsional frequencies of the system is essential\cite{abb_tnf,2}.

% This task can be  challenging for a drivetrain, and
% may even require consideration of the finite stiffness of flexible parts such as the motor bed and
% surrounding structures. Accurate modeling of the mechanical system is crucial for predicting
% and analyzing machines, as well as for predicting shaft torques, stresses, and shaft fatigue life
% expenditure following electrical disturbances\cite{2}. Therefore, the modeling of the mechanical
% system is a key focus in the development of an online diagnostics tool for medium voltage drives
% converters.

\begin{figure}[thpb]
      \centering 
 \includegraphics[scale=.5]{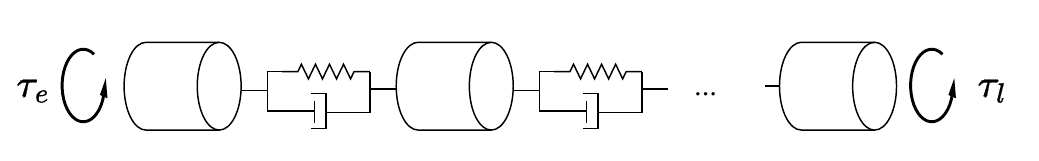}    
      \caption{ Lumped mass-spring model of a driveline.   }
      \label{fig:shaft}
   \end{figure}

The mechanical driveline of a variable speed industrial drive can be modeled as discrete inertias connected with inertia free
elastic elements (see Fig. \ref{fig:shaft}). 
The complex driveline system can be characterized by two transfer functions that describe the response of the shaft speed to the applied electrical torque and the load torque:
\begin{equation}
    \Omega(j\omega)=G_e(j\omega)\tau_e(j\omega)+G_l(j\omega)\tau_l(j\omega),
\end{equation}
where $\Omega$, $\tau_e$ and $\tau_l$ are, respectively, the discrete-Fourier transforms (DFT) the shaft speed, electrical and load torque signals. 

Similar to the previous approach, the system frequency response is approximated by the reduced order frequency response  around each resonant frequency:
 \begin{equation}
     G_e(j\omega)\approx\frac{1}{j\omega}\frac{c_e}{-\omega^2+2j\alpha \omega +\beta},
 \end{equation}
The term $1/j\omega$ is an integrator term that is explicitly taken out. The same assumption holds for $G_l$.

% Deriving the resonant frequencies is carried out by finding the resonating poles of $G_e$ (which are identical to poles of $G_l$). Around each resonant frequency, the behavior of the transfer function is mainly dominated by the complex-conjugate poles. That is, $G_e(j\omega)$ can be approximated by 
% \begin{equation}
%     G_e(j\omega)\approx\frac{1}{j\omega}\frac{c_e}{-\omega^2-2j\alpha \omega +\beta},
% \end{equation}
    
% where $c_e\in \mathbb{C}$ and constant, $\alpha$ and $\beta$ are respectively the real part and squared magnitude of the resonant poles, and $1/j\omega$ is an integrator term. The same assumption holds for $G_l$.

% The resonant frequency is then given by
% \begin{equation}
%     f_r=\frac{1}{2\pi}\sqrt{\beta-\alpha^2}.
% \end{equation}
% Plugging in the reduced order model of Gi’s, and splitting the signals into real and imaginary parts, lead to 
% Splitting the real and imaginary parts lead to
\begin{multline}
     j\omega(-\omega^2+2j\alpha \omega +\beta)(\Omega_r(\omega)+j\Omega_i(\omega))=\\
    (c_{er}+jc_{ei})(\tau_{er}(\omega)+j\tau_{ei}(\omega))+(c_{lr}+jc_{li})(\tau_{lr}(\omega)+j\tau_{li}(\omega))
    \label{eq:main2} 
\end{multline}

% where $\Omega_r$, $\Omega_i$, $\tau_e$ and $\tau_i$ are, respectively, the real and imaginary parts of the speed and the torque signals. 

Furthermore, we may assume $\tau_l(j\omega)$ to be smooth and constant in the neighborhood of the resonance. So the last expression can be simplified to a constant complex value, denoted by $d$ in the sequel. That is,
\begin{equation}
    (c_{lr}+jc_{li})(\tau_{lr}(\omega)+j\tau_{li}(\omega))\approx d_r+jd_i.
\end{equation}
Splitting the real and imaginary parts lead to 
% Eq. \ref{eq:main} can be split into real and imaginary parts and reformulated as follows
\begin{equation}
\underbrace{ \omega^3\begin{bmatrix} \Omega_i  \\ -\Omega_r \end{bmatrix}}_\text{$z(\omega)$}
   =
    \underbrace{\begin{bmatrix}
        2\omega^2\Omega_r & \omega\Omega_i & \tau_{er} & -\tau_{ei} & 1& 0\\
       2\omega^2\Omega_i & -\omega\Omega_r & \tau_{ei} & \tau_{er} & 0& 1
    \end{bmatrix}}_\text{$H(\omega)$}x,
\label{eq:zHx2}
\end{equation}
where $x$ is the parameter vector defined as 
\begin{equation*}
x:=\begin{bmatrix} \alpha& \beta& c_{er}& c_{ei} & d_r& d_i \end{bmatrix}^T.
  \end{equation*}

% The Eq. \ref{eq:zHx} is in the form of $z(\omega)=H(\omega)x$, where $z$ and $H$ both depend on the frequency, whereas $x$ is a constant parameter vector. Therefore, the parameter vector, $x$, can be estimated using the recursive least square (RLS) method which is computationally efficient. The RLS algorithm runs only around the resonant frequency (for example within 20\% bound of the initial guess), and recursively updates the estimated parameter $\hat{x}$  according to the following set of equations,
% \begin{equation}
%     \hat{x}_{n+1}=\hat{x}_n + K(z(\omega_n)-H(\omega_n)\hat{x}_n),
% \end{equation}
% where subscript $n$ captures the index of the DFT frequency points used for the estimation, and the matrix $K$ is the optimal gain matrix derived by the following equation
% \begin{equation}
%     K=PH^T(HPH^T+R)^{-1},
% \end{equation}
% where $R$ is the covariance matrix on the $z(\omega)$, and $P$ is the covariance matrix of the estimate given by,
% \begin{equation}
%     P=(I-KH)P(I-KH)^T+KRK^T.
% \end{equation}

\subsection{Simulation Results}
\begin{figure}[thpb]
      \centering
      
 \includegraphics[scale=.6]{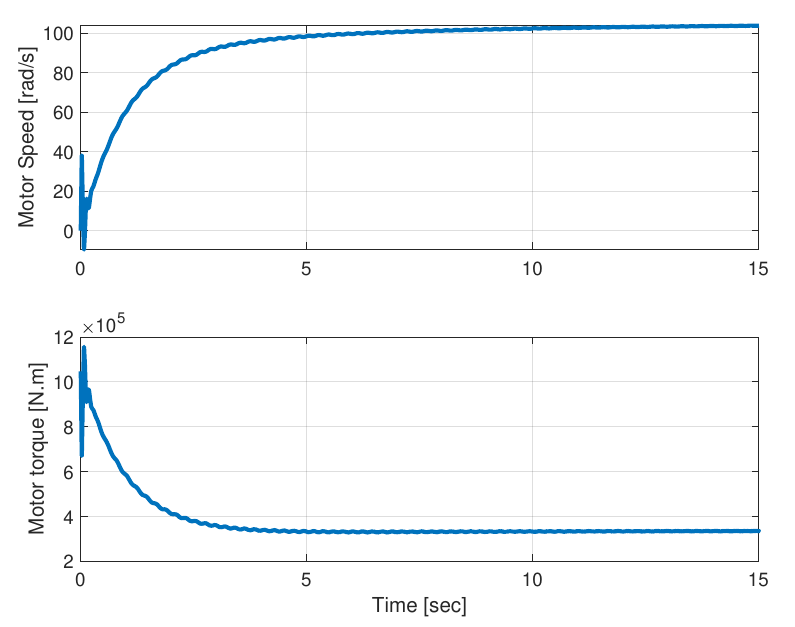}    
      \caption{Motor speed and torque trajectories.}
      \label{fig:speed}
   \end{figure}

\begin{figure}[thpb]
      \centering
      
 \includegraphics[scale=.6]{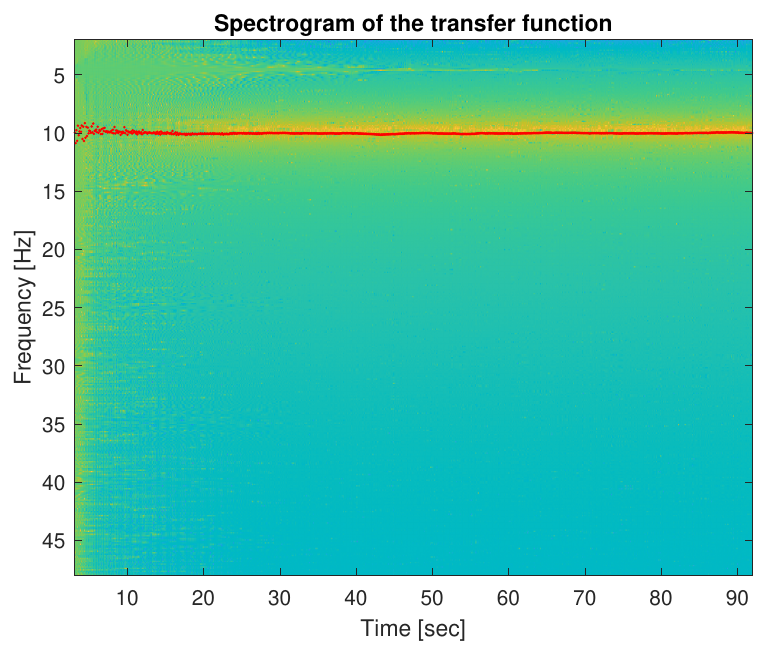}
           \caption{Spectrogram of the motor torque and speed signals. The red dots denote the estimate of the resonance frequency, $f_r$.  }
      \label{fig:spectogram}
   \end{figure}

\begin{figure}[thpb]
      \centering
      
 \includegraphics[scale=.6]{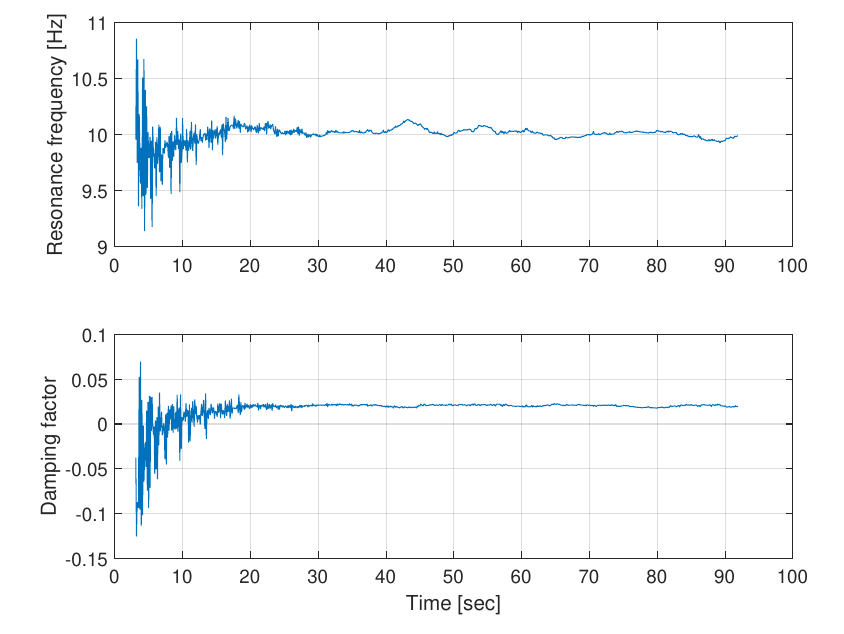}    
      \caption{ Estimate of the resonant frequency, $f_r$, and the damping factor, $\zeta$. }
      \label{fig:fr}
   \end{figure}

A mechanical driveline is simulated in Matlab/Simulink and the method described above is applied to the driveline signals to estimate a resonant mode. Fig. \ref{fig:speed} illustrates the motor speed and torque signals. We use the short-time Fourier transform (STFT) to convert the time-domain signals to the frequency-domain components. Fig. \ref{fig:spectogram} depicts the spectrogram of $\lvert G_e(\omega)\rvert$ during the transient and the steady-state.
 In this simulation, we only look at one resonance. Thus, the RLS algorithm sweeps from $f_{min}=8$Hz to $f_{max}=12$Hz, to detect the resonating frequency, which is initially known to be within this interval.  The resulting estimate is plotted in Fig. \ref{fig:spectogram} as red dots, and more clearly in Fig. \ref{fig:fr}. The actual value of $f_r$ is set to 9.8 Hz. The estimate of the damping factor is also plotted in Fig. \ref{fig:fr}.

\section{Summary}
\begin{figure}
  \centering
  \includegraphics[width=0.7\columnwidth]{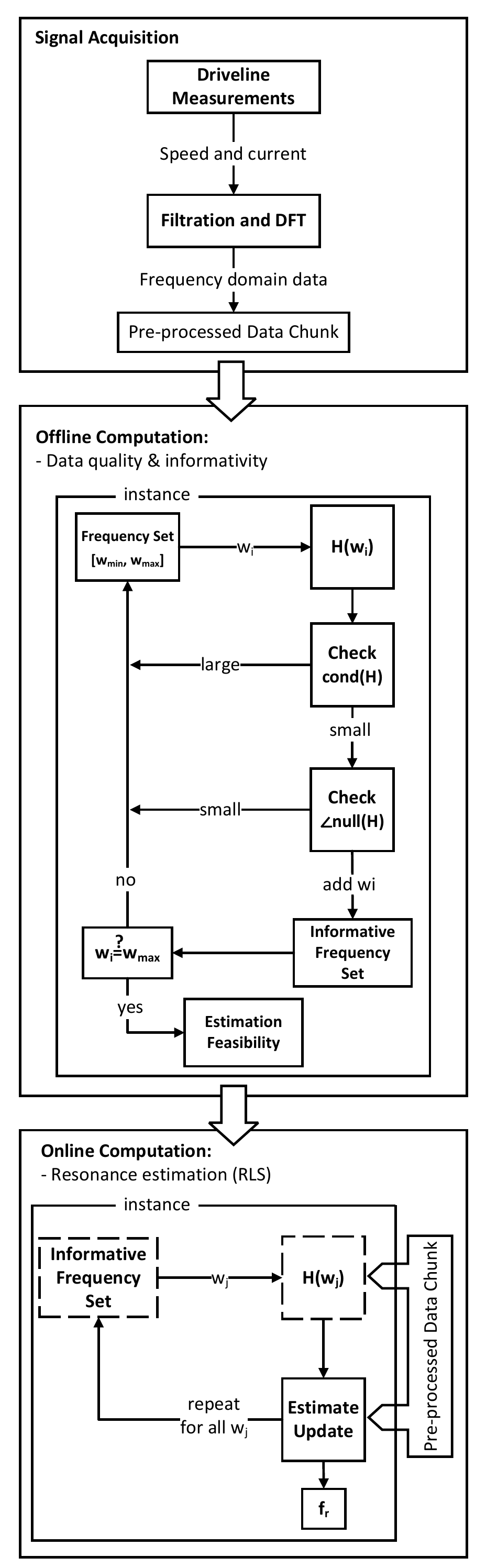}
     \caption{The resonance mode estimation procedure contains offline and online steps.    }
      \label{fig:flowchart}
\end{figure}

A novel solution is proposed for online estimation of the resonant modes. The method, described by a diagram in Fig. \ref{fig:flowchart}, is composed of an offline computation, where the data quality and informativity is analyzed\cite{ECC2024_imp,smi}, followed by an online efficient estimation process based on recursive least square (RLS), providing a computationally efficient solution. The RLS approach is applied to a narrower frequency range only around the resonant frequency, allowing the entire system response to be approximated by a second-order model. This simplification results in less computation for the resonance estimation, which is another novelty of the solution. All of the modules described can be run on an embedded system at a slower sampling rate than the controller or offline.

%%%%%%%%%%%%%%%%%%%%%%%%%%%%%%%%%%%%%%%%%%%%%%%%%%%%%%%%%%%%%%%%%%%%%%%%%%%%%%%%

%%%%%%%%%%%%%%%%%%%%%%%%%%%%%%%%%%%%%%%%%%%%%%%%%%%%%%%%%%%%%%%%%%%%%%%%%%%%%%%%

%%%%%%%%%%%%%%%%%%%%%%%%%%%%%%%%%%%%%%%%%%%%%%%%%%%%%%%%%%%%%%%%%%%%%%%%%%%%%%%%

% \section*{ACKNOWLEDGMENT}

% %%%%%%%%%%%%%%%%%%%%%%%%%%%%%%%%%%%%%%%%%%%%%%%%%%%%%%%%%%%%%%%%%%%%%%%%%%%%%%%%

% References are important to the reader; therefore, each citation must be complete and correct. If at all possible, references should be commonly available publications.\cite{IEEEexample:articledualmonths}

\bibliographystyle{IEEEtran}
\bibliography{resonance}

\end{document}